\documentstyle[12pt,bezier]{article}
\begin{document}
\def \inbar{\vrule height1.5ex width.4pt depth0pt}
\def \xC{\relax\hbox{\kern.25em$\inbar\kern-.3em{\rm C}$}}
\def \xR{\relax{\rm I\kern-.18em R}}
\newcommand{\xZ}{Z \hspace{-.08in}Z}
\newcommand{\xbe}{\begin{equation}}
\newcommand{\xee}{\end{equation}}
\newcommand{\xbea}{\begin{eqnarray}}
\newcommand{\xeea}{\end{eqnarray}}
\newcommand{\xnn}{\nonumber}
\newcommand{\xkt}{\rangle}
\newcommand{\xbr}{\langle}
\newcommand{\xlll}{\left( }
\newcommand{\xrrr}{\right)}
\newcommand{\xcun}{\mbox{\footnotesize${\cal N}$}}
\title{ Exact Semiclassical Evolutions in Relativistic and Nonrelativistic
Scalar Quantum Mechanics and Quantum Cosmology}
\author{Ali Mostafazadeh\thanks{E-mail: alimos@phys.ualberta.ca;
Fax: 1-(403) 492-0714}\\ \\
Theoretical Physics Institute, University of Alberta, \\
Edmonton, Alberta,  Canada T6G 2J1.}
\date{January 1997}
\maketitle

\begin{abstract}
The necessary and sufficient conditions for the exactness of the
semiclassical approximation for the solution of the Schr\"odinger
and Klein-Gordon equations are obtained. It is shown that the
existence of an exact semiclassical solution of the Schr\"odinger
equation determines both the semiclassical wave function and the
interaction potential uniquely up to the choice of the boundary
conditions. This result also holds for the Klein-Gordon equation. 
Its implications for the solution of the Wheeler-DeWitt equation for
the FRW scalar field minisuperspace models are discussed.
In particular,  exact semiclassical solutions of the Wheeler-DeWitt
equation for the case of massless scalar field and exponential
matter potentials are constructed. The existence of exact semiclassical
solutions for polynomial matter potentials of the form $\lambda\phi^{2p}$
is also analyzed. It is shown that for $p=1,~2$ and $3$, right-going
semiclassical solutions do not exist. A generalized semiclassical
perturbation expansion is also developed which is quite different
from the traditional $\hbar$ and $M_p^{-1}$-expansions. 
\end{abstract}
\vspace{.5cm}

\baselineskip=24pt

\newpage

\section{Introduction}

The semiclassical or WKB approximation  is usually discussed
in textbooks on nonrelativistic quantum mechanics  in the context
of stationary states, i.e., determination of the energy eigenvalues and
eigenfunctions, \cite{qm}. This approximation can also be used to
obtain approximate and in some cases
exact solutions of the dynamical problem, i.e., full Schr\"odinger equation.
To the best of my knowledge, however, the utility of the semiclassical
approximation in obtaining exact solutions of the Schr\"odinger equation
has not been fully explored. 

The same seems to be the case for the relativistic quantum mechanics.
The importance of the semiclassical approximation in the relativistic case
is probably best appreciated in quantum cosmology, \cite{page,wiltshire},
specifically, in the analysis of the Wheeler-DeWitt equation which is
essentially a Klein-Gordon equation on a superspace, \cite{dewitt}. 

In more general terms, the semiclassical approximation is usually
viewed as an approximation scheme in which one neglects all but
the first term in an asymptotic perturbation expansion of the solution
of a linear differential equation. Typical examples of such an
asymptotic expansion are the loop expansions of quantum mechanics
and field theory where the perturbation parameter is the Planck
constant\footnote{Here I have assumed that the kinetic term in the
Lagrangian does not involve a coupling constant. For example
in nonrelativistic quantum mechanics in a Euclidean space, this
corresponds to the case where the mass $m$ of the particle is set to
unity. Otherwise, the perturbation parameter for the loop expansion
is $\hbar/\sqrt{m}$.} $\hbar$, \cite{qm}. In the context of quantum cosmology
the relevant perturbation parameter is the gravitational coupling constant
(or inverse of the Planck mass $M_p$), \cite{singh,keifer,kim}. Usually,
these perturbation expansions are singular and it is difficult, if not
impossible, to obtain their precise structure.

There is a more universal alternative for defining the
semiclassical approximation where the validity of the approximation
is not linked with the values of the physical constants but determined
by the properties of the wave function. In this approach one uses the
polar representation of the wave function
	\xbe
	\psi({\bf x};t)=R({\bf x};t)\:e^{iS({\bf x};t)/\hbar}\;,
	\label{polar}
	\xee
and obtains two coupled nonlinear differential equations for the amplitude
$R$ and the phase (angle) $S$ of $\psi$ by substituting (\ref{polar}) in the
dynamical equation. As it is demonstrated for the Schr\"odinger and
Klein-Gordon equations in sections~2 and~3 below, there emerges a
quantity called the quantum potential $Q$ which controls the coupling
of these two equations.  In other words, if  $Q$ which depends only on $R$
happens to be negligible, then one of the equations decouples
from the other. The decoupled equation which only involves $S$ turns
out to satisfy a Hamilton-Jacobi equation. Thus, for $Q=0$, $S$ can be
identified with the classical action function of the corresponding
classical theory. This observation is originally due to Bohm \cite{d-bohm}.
It provides the basis for the de~Broglie-Bohm causal or ontological
interpretation of quantum mechanics, \cite{causal}. The latter has
recently been applied to problems of quantum cosmology by several
authors, \cite{db-qc}. 

The idea of the quantum potential leads to a precise criterion for
the validity of the semiclassical approximation, namely the condition
$Q\approx 0$. More precisely, one has the following
	\begin{itemize}
	\item[]{\bf Definition}:~ {\em A wave function is said to be
semiclassical if the corresponding quantum potential vanishes
identically.}
	\end{itemize}
Note that the quantum potential $Q$ is determined by the amplitude
$R$ of the wave function. Thus, the validity of the semiclassical
approximation has nothing to do with the value of the physical
constants which are fixed by nature. It is solely decided on the basis
of the particular form of the wave function. This in turn depends on the
interaction potential and the boundary conditions of the problem.

The purpose of this article is to derive the necessary and sufficient
conditions on the interaction potential and the boundary conditions
under which the dynamical equations, namely the Schr\"odinger
equation in the nonrelativistic case and Klein-Gordon equation in
the relativistic case, are exactly solved by a semiclassical wave
function. This is done in sections~2 and~3. Here the problem of
the classification of all potentials which allow for exact semiclassical
wave functions is solved. Section~4 includes a detailed analysis of
the $(1+1)$-dimensional Klein-Gordon equation. The results are then
applied in section~5 for the study of solutions of the Wheeler-DeWitt
equation for FRW scalar field minisuperspace models. Here several
exact semiclassical solutions are constructed. In section~6, the ideas
and the results of the preceding sections are used to develop a novel
semiclassical perturbation theory. The latter yields the semiclassical
approximation in the zero-th order of the perturbation theory. The
higher order corrections are shown to satisfy linear differential
equations with vanishing boundary conditions. In this way the
information about the boundary conditions of the original problem
is included in the zero-th (semiclassical) terms and the definition
of the perturbation potential. The proposed semiclassical perturbation
expansion is quite different from the traditional $\hbar$ and $M_p^{-1}$
expansions used in quantum mechanics and quantum cosmology.

\section{Nonrelativistic QM: Schr\"odinger Equation}

Consider the Schr\"odinger equation
	\xbea
	&&i\hbar\frac{d}{d t}\,\psi(t)=\hat H\psi(t)\;,~~~~\psi(0)=\psi_0
	\label{sch-eq}\\
	&&\hat H=\frac{1}{2m}\,\left[\hat{\bf p}-{\bf A}(\hat{\bf x};t)\right]^2+
	V(\hat {\bf x;}t)\;,\xnn
	\xeea
where $\psi$ is a state vector represented in the position representation
by the complex scalar wave function $\xbr {\bf x}|\psi(t)\xkt=\psi({\bf x};t)$,
${\bf A}$ is an electromagnetic vector potential, and $V$ is a scalar 
interaction potential. 

Inserting Eq.~(\ref{polar}) in the Schr\"odinger equation (\ref{sch-eq})
and making use of  $\xbr {\bf x}|\hat {\bf p}=-i\hbar{\bf\nabla}\xbr {\bf x}|$,
one obtains
	\xbea
	&&\partial_t S({\bf x};t)+H({\bf x},{\bf p}_*;t)+Q({\bf x};t)=0\;,
	\label{q-hj-eq}\\
	&& \partial_t\rho({\bf x};t)+{\bf \nabla}\cdot{\bf J}({\bf x};t)=0\;,
	\label{conti}
	\xeea
where $H=H({\bf x},{\bf p};t)$ is the classical Hamiltonian, $\rho:=R^2$,
${\bf p}_*:={\bf\nabla} S$, $Q:=-\hbar^2{\bf\nabla}^2 R/(2 m R)$ is
the quantum potential, and  ${\bf J}:=\rho {\bf v}_*$ with ${\bf v}_*:=
({\bf p}_*-{\bf A})/m$, is the probability current. 

Eq.~(\ref{q-hj-eq}) is the quantum analog of the Hamilton-Jacobi
equation
	\xbe
	\partial_t S({\bf x};t)+H({\bf x},{\bf p}_*;t)=0\;,
	\label{hj-eq}
	\xee
of the classical mechanics, \cite{goldstein}. It differs from the latter because
of  the presence of the quantum potential $Q$.  Eq.~(\ref{conti}) is the
continuity equation signifying the conservation of the probabilities.
	
According to the above definition, the semiclassical or WKB approximation
provides the exact solution of the Schr\"odinger equation, if and only if in
addition to Eqs.~(\ref{q-hj-eq}) and (\ref{conti}), one has
	\xbe
	Q:=\frac{-\hbar^2}{2m}\:\frac{{\bf\nabla}^2 R}{R}=0~~~
	\Longleftrightarrow~~~{\bf\nabla}^2 R=0\;,
	\label{condi}
	\xee
i.e., $R$ is a solution of the Laplace equation.
In this case, Eq.~(\ref{q-hj-eq}) reduces to the Hamilton-Jacobi
equation (\ref{hj-eq}). Therefore, the necessary and sufficient conditions
for the exactness of the semiclassical approximation are (\ref{hj-eq}),
(\ref{conti}), and (\ref{condi}). These equations may be viewed as three partial
differential equations for the three unknown functions $R$,  $S$
and $V$. 

Eq.~(\ref{condi}) does not involve time-derivatives. It is really a constraint
equation which can be independently solved.  Solving Eq.~(\ref{condi}) and
substituting the result in
(\ref{conti}), one finds a first order equation for ${\bf v}_*$ which in turn
yields ${\bf p}_*$. This leads to another first order differential equation
for $S$. The potential $V$ is then obtained by solving the latter equation
and substituting the result in Eq.~(\ref{hj-eq}).

There is an alternative way of solving the continuity equation (\ref{conti})
which involves writing it explicitly in terms of $S$, namely, considering
the solution of
	\xbe
	{\bf \nabla}\cdot (R^2{\bf \nabla} S)=R f\;,
	\label{conti-s0}
	\xee
where $f:=-2m\partial_t R+2{\bf\nabla}R\cdot{\bf A}+ R {\bf\nabla}
\cdot{\bf A}$. Now, let us define $\tilde S:=RS$. Then, it is an easy exercise
to show that $\tilde S$ is the solution of the following Poisson equation
	\xbe
	{\bf \nabla}^2\tilde S=f\;.
	\label{conti-s1}
	\xee
Here I have used in addition to Eq.~(\ref{conti-s0}) the constraint
equation (\ref{condi}). Hence, $S$ is given by a solution of the Poisson
equation (\ref{conti-s1}) divided by a (non-zero) solution of the Laplace
equation (\ref{condi}). Note that both of these equations are second
order elliptic differential equations with well-posed boundary value
problems.  Thus,  $R$, $S$ and consequently $V$ are uniquely
determined by the boundary conditions. This solves the problem of
the classification of all nonrelativistic (scalar) quantum systems with
an exact semiclassical solution for the Schr\"odinger equation by
relating the latter to the boundary conditions of the Laplace and Poisson
equations. It is also important to note that these boundary conditions
may in general depend on time which appears in the corresponding
equations as a parameter.

In order to demonstrate the utility of these findings in concrete terms, I
shall next consider the case where the classical configuration
space is one-dimensional. Here one can pursue according to
the former approach of integrating the continuity equation (\ref{conti})
by first solving for ${\bf v}_*$.

\subsection{One-Dimensional Configuration Spaces}

Consider a quantum system whose configuration space is the interval
$[x_1,x_2]\subset\xR$, and let the boundary conditions on the solution
of the Schr\"odinger equation (\ref{sch-eq}) be given by $\psi(x_1,t)=\psi_1(t)=:R_1(t)\exp[iS_1(t)/\hbar],~\psi(x_2,t)=\psi_2(t)
=:R_2(t)\exp[iS_2(t)/\hbar]$. In this case, Eq.~(\ref{condi}) becomes
$\partial_x^2 R=0$ which yields
	\xbe
	R=a(t)x+b(t)\;,
	\label{condi-1}
	\xee
where
	\xbe
	a(t)=\frac{R_1(t)-R_2(t)}{x_1-x_2}\:,~~~~~b(t)=
	\frac{x_1R_1(t)-x_2R_2(t)}{x_1-x_2}\;.
	\label{a-b}
	\xee
Substituting (\ref{condi-1}) in Eq.~(\ref{conti}) and integrating
the resulting differential equation, one obtains
	\xbea
	v_*&=&\frac{1}{(ax+b)^2}\left[-\frac{2}{3}a\partial_t a\: x^3-
	(a\partial_t b+b\partial_t a)x^2-2b\partial_t b\: x+c\right]\;,
	\label{v=}\\
	S&=&d +m\left[-(\frac{2a\partial_t a}{3})I_3
	-(a\partial_t b+b\partial_t a)I_2-2b\partial_t b\: I_1+c\: I_0\right]
	+\int A dx\;,
	\label{s=}
	\xeea
where $c=c(t)$ and $d=d(t)$ are functions of time determined by equating
the right hand side of (\ref{s=}) with $S_1$ and $S_2$ at $x=x_1$ and 
$x=x_2$, respectively, and
	\[I_k:=\int \frac{x^k dx}{(ax+b)^2}\;,~~~~k=0,1,2,3\;.\]
More explicitly, one has
	\xbea
	I_0&=&\frac{-1}{a(ax+b)}\:,~~~I_1\:=\:\frac{1}{a^2}\left[\ln|ax+b|+
	\frac{b}{ax+b}\right]\;,~~~
	I_2\:=\:\frac{1}{a^3}\left[ax+b-2b\ln|ax+b|+\frac{b^2}{ax+b}\right]\;,\xnn\\
	I_3&=&\frac{1}{a^4}\left[\frac{1}{2}(ax+b)^2-3b(ax+b)+3b^2\ln|ax+b|+
	\frac{b^3}{ax+b}\right]\;.\xnn
	\xeea
The potential is then obtained using Eq.~(\ref{hj-eq}), namely
	\xbe
	V=-\partial_t S-\frac{m}{2}v_*^2\;.
	\label{V=}
	\xee
It has the following general form:
	\[V=C_0(t)\ln[a(t)x+b(t)]+
	\frac{\sum_{\ell=0}^6 C_\ell(t)x^\ell}{[a(t)x+b(t)]^4}
	-\int  \partial_t A(x;t) dx\;,\]
where $C_\ell$, with $\ell=0,1,\cdots 6$, depend on $a$, $b$,
$c$, and $d$.

In view  of the above analysis, one can reach the following conclusions:
	\begin{itemize}
	\item[---] The condition of the exactness of the semiclassical
	approximation together with the boundary conditions determine
	both the semiclassical wave function and the potential uniquely;
	\item[---] For $x_1\to-\infty$ and $x_2\to\infty$, i.e., for a particle
	in $\xR$,  a smooth semiclassical wave function is not normalizable.
	It corresponds to a scattering state;
	\item[---] More general  exact semiclassical wave functions may
	be obtained by allowing a countable number of discontinuities.
	The effect of these discontinuities is to divide the interval $[x_1,x_2]$
	into a collection of subintervals in the interior of which the wave
	function and potential are given by the above expressions. The
	boundary conditions corresponding to each subinterval can be
	chosen freely. They determine the global structure of the wave
	function and the potential which can now be more complicated.
	This observation can also be used to devise an approximation
	scheme for the solution of the Schr\"odinger equation, by
	approximating the solution by a locally semiclassical one.
	\end{itemize}

Next, let us consider the following special cases:
	\begin{itemize}
	\item[1)] { {\em Constant Boundary Conditions}: $\partial_t\psi_1=
	\partial_t\psi_2=0$ with $R_1\neq R_2$}\\
	In this case, $a$ and
	$b$ do not depend on time. This simplifies the above formulae
	considerably. One has:
	\xbea
	v_*&=&\frac{c(t)}{(a x+b)^2}\;,~~~S\:=\:d(t)-\frac{m c(t)}{a(ax+b)}
	+\int A(x)dx\;,
	\label{S1=}\\
	V&=&-\partial_t d(t)+\frac{m \partial_tc(t)}{a(ax+b)}-
	\frac{mc(t)^2}{2(ax+b)^4}-\int  \partial_t A(x)dx\;,
	\label{V1=}
	\xeea
where
	\xbea
	c&=&\frac{(S_2-S_1)+(\gamma_1-\gamma_2)}{\zeta_1-\zeta_2}\:,~~~
	d\:=\:\frac{(\zeta_1S_2-\zeta_2 S_1)+(\zeta_2\gamma_1-\zeta_1\gamma_2)}{
	\zeta_1-\zeta_2}\:,\xnn\\
	\gamma_i&:=&\gamma(x_i)\;,~~~\gamma(x)\::=\: \int A(x)dx\;,~~~
	\zeta_i\::=\:\frac{m}{a(ax_i+b)}\:,~~~~~{\rm with}~~~i=1,2.\xnn
	\xeea
In particular, for  $A=0$, $\gamma_i$ vanish and $c$ and $d$ are
constant.  This leads to a further simplification of Eqs.~(\ref{S1=})
and (\ref{V1=}) and yields
	\[S=d-\frac{m c}{a(ax+b)}\;,~~~~
	V=-\frac{mv_*^2}{2}=-\frac{mc^2}{2(a x+b)^4}\;.\]
In this case both the potential and the action function turn out to
be time-independent. This corresponds to a semiclassical zero
energy eigenfunction.
	\item[2)] {{\em Amplitude-Periodic Boundary Conditions}, i.e.,
$R_1(t)=R_2(t)$} \\  In this case, $a=0$ and $R=b(t)$. 
Then, a similar analysis leads to:
	\xbea
	v_*&=&\frac{-2b\partial_t b\:x+c}{b^2}\;,~~~
	S\:=\:d+m\left[-(\partial_t \ln b)\:x^2+\frac{c}{b^2}\:x\right]+\int A\: dx\;,\xnn\\
	V&=&-\partial_t d-m\left[-(\partial_t^2\ln b) x^2+
	\partial_t(c/b^2)x\right]-\frac{m}{2}\left[\frac{-2b\partial_t b\:x+c}{b^2}\right]^2
	-\int\partial_t A\: dx
	\;.\xnn
	\xeea
Here, $c$ and $d$ depend on the phases $S_i$ of $\psi_i$ according to:	\[c=\frac{\Sigma_2(t)-\Sigma_1(t)}{\alpha_2(t)-\alpha_1(t)}\;,~~~
	d=\frac{\Sigma_1(t)\alpha_2(t)-\Sigma_2(t)\alpha_1(t)}{
	\alpha_2(t)-\alpha_1(t)}\;,\]
where
	\[\alpha_i:=\frac{mx_i}{b^2}\;,~~~\Sigma_i:=S_i-\gamma_i+
	m(\partial_t\ln b) \:x_i^2\;.\]
Note also that in this case, for $A=$ constant, the potential $V$ is at most a
quadratic polynomial in $x$. For, $b=R_1=R_2=$ constant, $V$ is either
a first order or a zero-th order polynomial in $x$. For example, for $b=1$,
$S_i=\omega_it$,  with $\omega_i$ being real constants, and $A=0$, one has:
	\xbea
	c&=&\frac{(\omega_2-\omega_1)t}{m(x_2-x_1)}\;,~~~
	d\:=\:\frac{(\omega_1x_2-\omega_2x_1)t}{x_2-x_1}\;,\xnn\\
	S&=&(\frac{t}{x_2-x_1})[(\omega_1x_2-\omega_2x_1)+
	(\omega_2-\omega_1)x]\;,\xnn\\
	V&=&(\frac{-1}{x_2-x_1})[(\omega_1x_2-\omega_2x_1)+
	(\omega_2-\omega_1)x]-\frac{1}{2m}
	\left[\frac{(\omega_2-\omega_1)t}{(x_2-x_1)}\right]^2
	\;.\xnn
	\xeea
Another interesting case is when $b=e^{\omega t/2}$, for some constant $\omega$. 
Then, the potential is a quadratic polynomial in $x$ with the coefficient
of the quadratic term being a constant, namely, $-m\omega^2/2$. In
the latter case if the phases $S_1$ and $S_2$ and the vector potential
$A$ are also time-independent, then so are $c$ and $d$. Therefore, one has
a time-independent quadratic potential, namely
	\[V=-\frac{m}{2}(\omega x+c)^2\;.\]
	\end{itemize}

\subsection{Multi-Dimensional Configuration Spaces}
	For an $n$-dimensional configuration space, with $n>1$, the classification
of the exact semiclassical wave functions and the corresponding potentials
is more involved.  This is mainly because in this case the constraint
equation (\ref{condi}) is the $n$-dimensional Laplace equation
${\bf \nabla}^2R=0$. 

In Cartesian coordinates, one can use the method of separation of variables
to express $R$ as a sum of the basic solutions
	\xbe
	\prod_{i=1}^{n}\left\{a_{i}(t){\cal S}
	[\kappa_i(t)x^i]+b_{i}(t){\cal C}[\kappa_i(t)x^i]\right\}\;.
	\label{R-mult}
	\xee
where $(x^1,\cdots,x^n):={\bf x}$, $\kappa_i$, $a_{i}$,  and
$b_{i}$ are real functions of time which are determined by the
boundary conditions on the wave function, $\kappa_i$ satisfy 
	\xbe
	\sum_{i=1}^n \eta_i\:\kappa_i^2(t)=0\;,
	\label{k-condi}
	\xee	
with $\eta_i=\pm1$, and ${\cal S}$ (resp.\  ${\cal C}$) stands for either of
$\sin$ or $\sinh$ (resp.\ $\cos$ or $\cosh$) depending on whether
$\eta_i=-1$ or $+1$, respectively. Clearly, one can choose one of the
$\eta_i$'s positive and others negative.

Having found the expression for $R$, one then proceeds by
integrating the continuity equation (\ref{conti}) which yields $S$.  In
view of the above analysis, $S=\tilde S/R$, where $\tilde S$ is
a solution of the Poisson equation (\ref{conti-s1}). Using the well-known
Green's function methods \cite{jackson} of solving the Poisson equation,
one may express $S$ in an integral form. For example, if the configuration
space is $\xR^3$, then
	\xbe
	S({\bf x};t)=\frac{1}{ R({\bf x};t)}\left\{ \tilde S_0({\bf x};t)+
	\frac{1}{2\pi}\int dx^{'3}~\left[\frac{
	m\partial_t R({\bf x'};t)-{\bf\nabla}R({\bf x'};t)\cdot{\bf A}({\bf x'};t)-
	\frac{R({\bf x'};t)}{2} {\bf\nabla}
	\cdot{\bf A}({\bf x'};t)}{|{\bf x}-{\bf x'}|}\right]\right\}\;,
	\label{S=int}
	\xee
where $\tilde S_0$ is a solution of the Laplace equation determined
by the boundary conditions.

Similarly to the one-dimensional case, in the case that the configuration
space is $\xR^n$, a smooth semiclassical wave function cannot be
normalized. More generally, it cannot vanish at infinity, nor can it be
localized. This is a direct consequence of Eq.~(\ref{k-condi}).

The main difference with the one-dimensional case is that here one
has a much richer structure as far as the general form of the wave function
and the potential is concerned. Unfortunately,
since without knowing the specific form of  the boundary conditions one
cannot express $R$ and  $S$ in a closed form, an explicit classification
of the semiclassical wave functions and the corresponding potentials
for $n>1$ is not available. Nevertheless, it is evident that by choosing
the boundary conditions appropriately one can obtain a large variety of
potentials. 

In order to demonstrate the validity of this claim, I shall next concentrate
on the special cases  where the amplitude of the semiclassical
wave function is independent of ${\bf x}$. A simple example of this
is a particle in a cubical cavity of side length $L$ with the boundary
conditions:
	\[\left.\psi\right|_\partial=N(t)\:e^{iS_\partial({\bf x};t)/\hbar}\;,\]
where the symbol $\partial$ stands for the boundary of the cavity.
In this case, $R=N(t)$ is the unique solution of the constraint equation
(\ref{condi}), and ${\bf\nabla}R=0$. This reduces Eq.~(\ref{conti-s0}) to
the simple Poisson equation
	\xbe
	{\bf \nabla}^2 S=-2m\partial_t\ln N\;,
	\label{po-eq}
	\xee
where I have chosen the Coulomb gauge so that ${\bf\nabla}\cdot{\bf A}=0$.
Note that the source term on the right hand side of Eq.~(\ref{po-eq}) only
depends on time. Hence, one can define $\check S:=S+m[\partial_t\ln N]\: |{\bf x}|^2$
and reduce this equation to the Laplace equation
	\xbe
	{\bf \nabla}^2 \check S=0\:.
	\label{laplace}
	\xee
Since the set of solutions of the Laplace equation are in one to one
correspondence with the set of boundary conditions which is
a very large function space, one obtains a large class of potentials.

Next, consider the following simple subcases.
	\begin{itemize}
	\item[1)] $\check S_\partial=\left.{\bf {\cal K}}(t)\cdot{\bf x}\right|_\partial$, 
where ${\bf {\cal K}}$ is an ${\bf x}$-independent vector-valued function
of time. Then, $\check S={\bf{\cal K}}(t)\cdot{\bf x}$ clearly satisfies the Laplace
equation (\ref{laplace}) and one has:
	\xbea
	S&=&{\bf{\cal K}}(t)\cdot{\bf x}-m[\partial_t\ln N(t)]\: |{\bf x}|^2\xnn\\
	V&=&m[\partial_t^2\ln N(t)]|{\bf x}|^2-\partial_t{\bf {\cal K}}(t)\cdot{\bf x}-
	\frac{1}{2m}|2m[\partial_t\ln N(t)]{\bf x}+{\bf A}({\bf x};t)-{\bf {\cal K}}(t)|^2\;.
	\xnn
	\xeea
This case is the multi-dimensional analog of example~2 of section~2.1.
	\item[2)] Consider the case $n=2$,  i.e., a square with boundaries
$x^1=:x=0,L$ and $x^2=:y=0,L$, and boundary conditions:
$\check S=0$ for $x=0,L,~y=0$, and $\check S=\alpha(t)\sin(\pi x/L)$ for $y=L$. Then,
one can easily show that the solution of the Laplace equation (\ref{laplace})
is given by
	\[\check S=\frac{\alpha(t)\sin(\pi x/L)\sinh(\pi y/L)}{\sinh\pi}\:.\]
Hence, one has
	\xbea
	S&=&\frac{\alpha(t)\sin(\pi x/L)\sinh(\pi y/L)}{\sinh\pi}-
	m[\partial_t\ln N(t)]\: |{\bf x}|^2\xnn\\
	V&=&m[\partial_t^2\ln N(t)] |{\bf x}|^2-\frac{\partial_t\alpha(t)
	\sin(\pi x/L)\sinh(\pi y/L)}{\sinh\pi}-\xnn\\
	&&\frac{\alpha^2(t)}{2m\sinh^2\pi}\left\{
	\left[ \frac{\pi}{L}\cos(\pi x/L)\sinh(\pi y/L)-2m[\partial_t\ln N(t)]x-
	\frac{\sinh(\pi)A_x}{\alpha(t)}\right]^2+\right.\xnn\\
	&&\left.\left[ \frac{\pi}{L}\sin(\pi x/L)\cosh(\pi y/L)-2m[\partial_t\ln N(t)]y-
	\frac{\sinh(\pi)A_y}{\alpha(t)}\right]^2\right\}\;.\xnn
	\xeea
These relations show that unlike the one-dimensional case, here
the wave function and the potential can be quite complicated.
	\end{itemize}

So far, I have only considered  cases for which  ${\bf\nabla}R=0$.
This is precisely the condition demanded in the traditional semiclassical
approximation. It is usually argued that the semiclassical approximation is 
valid if the amplitude $R$ of the wave function is a slowly varying function
of ${\bf x}$. As it is clear from the above analysis, this is only a sufficient
condition, not a necessary one.  There is an infinite number of examples
where $R$ is a rapidly changing function of ${\bf x}$ but the semiclassical
approximation is not only valid, but it yields the exact result. Specific
examples can be constructed by simply choosing $R$ to be a rapidly
changing solution of the Laplace equation. For instance consider a quantum
system with the geometry of the preceding example, but the boundary
conditions which lead to
	\begin{itemize}
	\item[3)]  $R(x,y;t)=R(x,y)=N_0\sin(\ell \pi x/L)\sinh(\ell\pi x/L)$ and
	$S(x,y;t)=S(x,y)={\bf{\cal K}}\cdot{\bf x}/R$, where $N_0$ and  $\ell$
	are real and integer constants, respectively, and ${\bf{\cal K}}$ is a
	constant vector. One can easily check that $R$ and $S$ satisfy
	Eqs.~(\ref{condi}) and (\ref{conti-s0}). However, 
	\[ |{\bf\nabla} R|=|\frac{\ell N\pi}{L}\cos(\frac{\ell  \pi x}{L})\cosh(\frac{\ell  \pi y}{L})|
	\sqrt{\tan^2(\frac{\ell  \pi x}{L})+\tanh^2(\frac{\ell  \pi y}{L})}\]
	can be made arbitrarily large by choosing  large values for $\ell$,
	i.e., $R$ is not a slowly varying function of $x$ and $y$. Note also
	that in this case both $R$ and $S$ are time-independent. Thus, the
	wave function describes a zero-energy eigenfunction of the
	time-independent potential
	\[V=-\frac{({\bf\nabla}S)^2}{2m}=
	\frac{-1}{2mR^2}\left[{\bf{\cal K}}^2+
	\frac{({\bf{\cal K}}\cdot{\bf x})^2({\bf\nabla}R)^2}{
	R^2}-\frac{2({\bf{\cal K}}\cdot{\bf x})({\bf{\cal K}}
	\cdot{\bf \nabla}R)}{R}\right]\;.\]
	\end{itemize}
This example clearly shows how the present analysis generalizes the
results of the traditional semiclassical approach to quantum mechanics.

\section{Relativistic QM: Klein-Gordon Equation}

Consider the Klein-Gordon equation
	\xbe
	\left[(\partial^\mu-A^\mu)(\partial_\mu-A_\mu)-V(x)\right]\psi(x)=0\;,
	\label{kg-eq}
	\xee
where $A_\mu$ are components of an electromagnetic gauge field, $V$
is a scalar interaction potential (including the mass term in the massive case),
and $x$ stands for the four vector $(x^\mu)$. In the following, I shall follow 
the relativists' convention for the Minkowski metric, namely, $(\eta_{\mu\nu})
={\rm diag}(-1,1,\cdots,1)$, and set $c=\hbar=1$.

In the polar representation (\ref{polar}), the Klein-Gordon equation
is written as
	\xbea
	&&(\partial^\mu S-A^\mu)(\partial_\mu S-A_\mu)+V+Q=0\;,
	\label{q-hj-eq/kg}\\
	&&\partial_\mu J^\mu=0\;,
	\label{conti/kg}
	\xeea
where $Q:=-\partial^\mu\partial_\mu R/R$ is the quantum potential
and $J^\mu:=\rho(\partial^\mu S-A^\mu)$, with $\rho:=R^2$, is the conserved
current. Again, Eq.~(\ref{q-hj-eq/kg}) is the quantum analog of the
Hamilton-Jacobi equation for a classical relativistic particle
	\xbe
	(\partial^\mu S-A^\mu)(\partial_\mu S-A_\mu)+V=0\;,
	\label{hj-eq/kg}
	\xee
and Eq.~(\ref{conti/kg}) is the continuity equation associated with the
charge conservation.

A semiclassical Klein-Gordon field is defined by the condition	\xbe
	Q:=-\frac{\partial^\mu\partial_\mu R}{R}=0~~~~
	\Longleftrightarrow~~~~\partial^\mu\partial_\mu R=0\;.
	\label{condi/kg}
	\xee
Therefore, the semiclassical or WKB approximation is exact, if and only if
the relations (\ref{condi/kg}), (\ref{hj-eq/kg}), 
and (\ref{conti/kg}) are satisfied. As in the nonrelativistic case, these
three equations may be used to determine the three unknown real functions
$R$, $S$, and $V$. This is done by first solving Eq.~(\ref{condi/kg}) which is
already decoupled from the other two. This is a wave equation for $R$. Its
general solution is given by a linear combination of the functions of the
form
	\xbe
	W_{\hat{\bf k}}=W_{\hat{\bf k}}(x^0-{\bf x}\cdot\hat{\bf k})	\label{packet}
	\xee
where $x=(x^0,{\bf x})$ belongs to the $(n+1)$-dimensional Minkowski
space ${\cal M}^{n+1}$ or a subset of  ${\cal M}^{n+1}$, and $\hat{\bf k}$
is a unit $n$-vector defining the null wave $(n+1)$-vector $k=k_0(1,\hat{\bf k})$.
A simple choice for $W_{\hat{\bf k}}$ which is essentially motivated by
the Fourier analysis of the wave equation is the plane waves
$\exp i(k_0x^0-{\bf x}\cdot{\bf k})$. Once $W_{\hat{\bf k}}$ are
chosen then the solution of Eq.~(\ref {condi/kg}) reduces to  the determination
of the coefficients of $W_{\hat{\bf k}}$ in the expansion of $R$. 

Next, one substitutes the expression for $R$  in the continuity
equation (\ref{conti/kg}) and integrates the resulting equation. The
basic strategy is similar to the nonrelativistic case.  In terms of $S$
the continuity equation (\ref{conti/kg}) takes the form
	\xbe
	\partial^\mu(R^2\partial_\mu S)=RF\;,
	\label{conti/kg-S}
	\xee
where $F:=2\partial_\mu R A^\mu+R\partial_\mu A^\mu$. Eq.~(\ref{conti/kg-S})
is the analog of  Eq.~(\ref{conti-s0}). It can further  be simplified by defining
$\tilde S:=RS$. This together with Eqs.~(\ref{condi/kg}) and (\ref{conti/kg-S})
leads to
	\xbe
	\partial^\mu\partial_\mu\tilde S=F\;,
	\label{conti/kg-1}
	\xee
i.e., $S=\tilde S/R$, where $\tilde S$ is a solution of the inhomogeneous
wave equation (\ref{conti/kg-1}). Having found $R$ and $S$, one can
use the Hamilton-Jacobi equation (\ref{hj-eq/kg}) to determine the form
of the potential.

There is a very important difference between the relativistic and
nonrelativistic cases.  Here the condition of the exactness of
the semiclassical approximation leads to two second order hyperbolic
equations, namely (\ref{condi/kg}) and (\ref{conti/kg-1}), whereas
in the nonrelativistic case one has two elliptic equations. One knows
from the theory of  hyperbolic differential equations that the
boundary-value problem for such equations is not  generally well-posed, i.e.,
for arbitrary boundary conditions, a solution may or may not exist and if
it does, it may not be unique. The well-posed problem for a hyperbolic
equation such as the wave equations (\ref{condi/kg}) and (\ref{conti/kg-1})
is the initial-value problem. In general for given initial data on a Cauchy
hypersurface, one can solve these equations and determine $R$, $S$,
and $V$ uniquely. One of the consequences of the hyperbolicity of
(\ref{condi/kg}) is that unlike the nonrelativistic case, here $R$ and
therefore the semiclassical wave function can be localized. In particular,
one can form a coherent wave packet which approximates the behavior
of a classical particle.

Restricting to the case where $R$ is a constant and adopting the
Lorentz gauge $\partial^\mu A_\mu=0$, one can reduce
Eq.~(\ref{conti/kg-S}) to a (homogeneous) wave equation for $S$:
	\xbe
	\partial^\mu \partial_\mu S=0\;,
	\label{wave-s}
	\xee
The general solution of this equation is also given as a linear combination
of  functions of the form (\ref{packet}). This is sufficient to conclude that even
for this special case $S$ and consequently $V$ can be quite complicated.
This shows that there is a large class of potentials which allow exact
semiclassical solutions of the Klein-Gordon equation. These potentials and the
corresponding semiclassical Klein-Gordon fields depend in a 
crucial way on the boundary conditions\footnote{Here
and in what follows, ``boundary conditions'' means ``initial'', ``boundary'', or
``mixed initial-boundary conditions'' for which there exists at least one solution.}.
This is especially important in quantum cosmology where there is an
ongoing controversy regarding the choice of the boundary conditions
for the wave function of the universe and also the form of the potential
in the Wheeler-DeWitt equation. In particular, for the FRW scalar field
minisuperspace models \cite{page,wiltshire}, the Wheeler-DeWitt
equation is precisely a $(1+1)$-dimensional Klein-Gordon equation
of the form (\ref{kg-eq}). For these models the form of the potential is
directly linked with the phenomenon of inflation \cite{inflation,page,wiltshire}.
On the other hand, most if not all the physical predictions which one hopes
to derive from such models are relevant to the regions of the minisuperspace
where the wave function is semiclassical. The results of this paper indicate that
at least one can rule out the cases where the existence of a semiclassical
solution (for some region of the minisuperspace) is inconsistent with
the form of the potential (in that region). This together with the requirements
imposed by inflation may be helpful in improving our understanding of
quantum cosmology. This motivates a closer analysis of the Klein-Gordon
equation in the $(1+1)$-dimensional Minkowski space.

\section{Klein-Gordon Equation in $(1+1)$-dimensions}

If the configuration space is the $(1+1)$-dimensional Minkowski 
space ${\cal M}^2$, then the general solution of the wave equations
(\ref{condi/kg}) and (\ref{conti/kg-1}) can be written in terms of four
real-valued functions $R_\pm$ and $\tilde S_\pm$, \cite{m-f},
	\xbea
	R(x,t)&=&R_+(x+t)+R_-(x-t)\;,
	\label{1-1-R}\\
	\tilde S(x,t)&=& \tilde S_+(x+t)+
	\tilde S_-(x-t)+\int dx' dt'G(x,t;x',t')F(x',t')\;,\xnn
	\xeea
where $F$ is the same as the one appearing in Eq.~(\ref{conti/kg-S})
and $G$ is the appropriate Green's function for the one-dimensional wave
equation, \cite{m-f}. The latter can be constructed out  of the advanced
and retarded Green's functions given by $G^\pm(x,t;x',t'):=[\theta(|x-x'
|\pm(t-t'))-1]/2$, where $+$ and $-$ label the advanced and
retarded Green's functions, respectively, and $\theta$ is the step function,
$\theta(z)=1$, if $z>0$; $\theta(z)=0$, if  $z<0$. The usual choice in typical
physical applications is the retarded Green's function $G^-$ which marks
a particular direction of time. Note that if the electromagnetic potential
$A_\mu$ is absent, $F=0$, the last term in Eq.~(\ref{1-1-S}) drops, and
there is no need for a Green's function, in particular, a direction
of time. Finally in view of $S=\tilde S/R$ one has	\xbe
	S(x,t)=\frac{1}{R_+(x+t)+R_-(x-t)}\left[ \tilde S_+(x+t)+
	\tilde S_-(x-t)+\int dx' dt'G(x,t;x',t')F(x',t')\right]\;.
	\label{1-1-S}
	\xee

Eqs.~(\ref{1-1-R}) and (\ref{1-1-S}) together with Eq.~(\ref{hj-eq/kg})
show that in general the exact semiclassical wave functions and the
corresponding  potentials are classified by the set ${\cal C}^4:=
\{(R_\pm,\tilde S_\pm)\}$, where ${\cal C}$ is the set of real-valued
functions of a single real variable.

Let us next concentrate on the case where there is no electromagnetic
interaction. Then, in view of Eq.~(\ref{hj-eq/kg}) the potential has
the general form
	\xbea
	V&=&\frac{-4[\tilde S'_+\tilde S'_-+ S^2R'_+R'_-
	-S(R'_+\tilde S'_-+R'_-\tilde S'_+)]}{(R_++R_-)^2}\;,\xnn\\
	&=&-4\left(\frac{\tilde S'_+}{\tilde S_++\tilde S_-}-
				\frac{R'_+}{R_++R_-}\right)
			\left(\frac{\tilde S'_-}{\tilde S_++\tilde S_-}-
				\frac{R'_-}{R_++R_-}\right)S^2
	\:=:\:{\cal F}[R_\pm,\tilde S_\pm]\;,
	\label{1-1-V}
	\xeea
where $R_\pm=R_\pm(x\pm t)$, $\tilde S_\pm=\tilde S_\pm(x\pm t)$,
a prime means first derivative of the corresponding function,
	\xbe
	 S=\frac{\tilde S_++\tilde S_-}{R_++R_-}\;,
	\label{1-1-S-0}
	\xee
and the function(al) ${\cal F}$ is defined for future use.
Eq.~(\ref{1-1-V}) is obtained by substituting (\ref{1-1-R}) and
(\ref{1-1-S-0}) in the Hamilton-Jacobi equation~(\ref{hj-eq/kg}).

Next, consider the simple case $R_\pm=1/2$. Then, $S=\tilde S_+
+\tilde S_-$ and $V=-4\tilde S'_+\tilde S'_-$. In particular, one has
the following interesting examples:
	\begin{itemize}
	\item[1)] { $S$ is  linear in time $t$:  $\tilde S_\pm=\omega_\pm(x\pm 
t)/2$ for some constants $\omega_\pm$}\\
In this case, $V=-\omega_+\omega_-=$ constant. This includes the case of a
free Klein-Gordon field of mass $\mu=\sqrt{-\omega_+\omega_-}$, since
in this case $V=\mu^2$. 
	\item[2)] { $S$ is quadratic in $t$ and $x$:  $\tilde S_\pm=\nu_\pm
	(x\pm t)^2/2$ for some constants $\nu_\pm$}\\
In this case, one obtains a quadratic potential of the form
$V=4\nu_+\nu_-(t^2-x^2)$. This corresponds to a Klein-Gordon
field with the time-dependent mass $\mu=2\sqrt{\nu_+\nu_-}~t$ and
quadratic interaction potential. The nonrelativistic limit of this case
is a time-dependent harmonic oscillator with imaginary frequency.
	\item[3)] { $S$ is a linear combination of exponential functions:
	$\tilde S_\pm=\nu_\pm e^{\omega_\pm(x\pm t)}$}\\
In this case, one has an exponential potential,  $V=V_0 e^{(\omega_++\omega_-)x+(\omega_+-\omega_-)t}$, where $V_0:=-4\omega_+\omega_-\nu_+\nu_-$. Clearly, by choosing
$\omega_-=\pm\omega_+=:\pm\omega$, one obtains exponential
potentials which depend only on $t$ or $x$, namely $V=V_0e^{2\omega x}$
and $V=V_0e^{2\omega t}$, respectively.
	\end{itemize}

These examples can also be described in the framework of the traditional
semiclassical approach, since $R$ is chosen to be unity. Similarly to the
nonrelativistic case, in order to demonstrate the generality of the present
analysis, one must consider the cases where $R$ is a rapidly
changing function of  $t$ and $x$, but the semiclassical approximation is 
nevertheless exact. Again typical examples can be constructed starting
from a rapidly changing solution of the wave equation (\ref{condi/kg}).
For instance, consider the case
	\begin{itemize}
	\item[4)]
	{ $R=R_0 e^{\omega(x-t)}$ and $S=S_0e^{-2\omega x}$}\\
Then one can check that Eqs.~(\ref{condi/kg}) and (\ref{conti/kg-S})
are satisfied and the corresponding semiclassical wave function is an
exact solution of the Klein-Gordon equation defined by the
time-independent potential $V=-4\omega^2S_0^2e^{-4\omega x}$.
	\end{itemize}

Eq.~(\ref{1-1-V}) provides a classification of the potentials for which
an exact semiclassical solution of the $(1+1)$-dimensional Klein-Gordon
equation, with $A_\mu=0$, exists. In practice, however, it is the potential
which is given, not the wave function. Hence, a more interesting question
is {\em whether for a given potential $V=V(x,t)$ there is a set of boundary
conditions which yields an exact solution of the Klein-Gordon equation.} 
One can alternatively ask whether a given potential belongs to the image
of the function ${\cal F}$ defined in Eq.~(\ref{1-1-V}).

In order to answer these questions, I shall first consider a special class
of boundary conditions, namely $R_+=0$. For this class one can
show that  ${\cal F}$ is not onto and the set of potentials of the from
Eq.~(\ref{1-1-V}), with $R_+=0$, forms a small subset of all possible
potentials. To see this, let us first substitute $R_+=0$ in (\ref{1-1-V}).
The resulting equation may then be viewed as a differential equation
for $\tilde S_-$, while $R_-$ and $\tilde S_+$ are treated as undetermined
functions. This leads to
	\xbe
	\tilde S'_-+(-\frac{R'_-}{R_-})\tilde S_-+\frac{R_-^2}{4}
	\left[\frac{V}{\tilde S'_+}-\frac{4R'_-\tilde S_+}{R_-^3}\right]=0\;,
	\label{s-=}
	\xee
which is a consistent first order ordinary linear differential equation for
$\tilde S_-$ provided that the bracket on its left-hand side depends
only on $x-t$. This puts an strong restriction on the form of the
allowed potentials. Namely,  the potential must be of the form:
	\xbe
	 V(x,t)={\cal X}(x-t)\tilde S_+'(x+t)+{\cal Y}(x-t)
	\tilde S_+(x+t)\tilde S_+'(x+t)\;,
	\label{V-s-=}
	\xee
where ${\cal X}$ is an arbitrary real-valued function and ${\cal Y}:=
4R'_-/R^3_-$. 

	\begin{itemize}
\item[] {\bf Proposition~1:}  {\em Let  $u:=x-t$ and $v:=x+t$ be null
coordinates in ${\cal M}^2$ and  $V:{\cal M}^2\to\xR$ be an analytic
function at $(0,0)\in{\cal M}^2$. Then, a necessary condition for $V$ 
to satisfy Eq.~(\ref{V-s-=}), for some functions ${\cal X}:\xR\to\xR$, 
${\cal Y}:\xR\to\xR$, and $\tilde S_+:\xR\to\xR$, is that  the
coefficients $V_{jn}$ in the power series expansion $\sum_{j,n=0}^\infty
V_{jn}u^jv^n$ of $V$ must satisfy one of the following two
relations}
	\xbea
	V_{jn}&=&\frac{
	(V_{k_1m_1}V_{jm_1}-V_{k_1m_2}V_{jm_2})V_{k_2n}+
	(V_{k_2m_2}V_{jm_2}-V_{k_2m_1}V_{jm_1})V_{k_1n}}{
	V_{k_1m_1}V_{k_2m_2}-V_{k_1m_2}V_{k_2m_1}}\;,
	\label{vjn-condi-1}\\
	V_{jn}&=&\frac{V_{jm_1}V_{k_1n}}{V_{k_1m_1}}\;,
	\label{vjn-condi-2}
	\xeea
with $(j,k_1,k_2)$ and $(n,m_1,m_2)$ being triplets of mutually
different arbitrary non-negative integers.
\item[] {\bf Proof:}  Substitute the power series expansions
	\xbea
	V&=&V(u,v)\:=:\:\sum_{j,n=0}^\infty
	V_{jn}u^jv^n\;,~~~~~~{\cal X}\:=
	\:{\cal X}(u)\:=:\:\sum_{j=0}^\infty {\cal X}_j u^j\;,
	\label{v-exp}\\
	{\cal Y}&=&{\cal Y}(u)\:=:\: \sum_{j=0}^\infty {\cal Y}_j u^j\;,~~~~~~
	\tilde S_+\:=\:\tilde S_+(v)\:=:\: \sum_{n=0}^\infty S_nv^n\;,
	\label{s-exp}
	\xeea
in Eq.~(\ref{V-s-=}). This leads to
	\xbe
	V_{jn}=(n+1)S_{n+1}{\cal X}_j+{\cal Y}_j\sum_{m=0}^n
	(n-m+1)S_mS_{n-m_1}=(n+1)S_{n+1}{\cal Z}_j+
	{\cal Y}_j\sum_{m=0}^{n-1}(n-m)S_{m+1}S_{n-m}\;,
	\label{Vjn}
	\xee
where ${\cal Z}_j:={\cal X}_j+S_0{\cal Y}_j$. Next, solve for the
sum on the right-hand side of (\ref{Vjn}). The result is
	\[\sum_{m=0}^{n-1}(n-m)S_{m+1}S_{n-m}=\frac{1}{{\cal Y}_j}\left[
	V_{jn}-(n+1)S_{n+1}{\cal Z}_j\right]\;.\]
Clearly the left-hand side of this equation is independent of $j$.
Hence, its right-hand side must also be independent of  $j$. Writing
the right-hand side for two different values of $j$ and equating the
results, one has
	\xbe
	 S_{n}=\frac{1}{n}\left( \frac{{\cal X}_j}{{\cal Y}_j}-
	\frac{{\cal X}_k}{{\cal Y}_k}\right)^{-1}\left(
	\frac{V_{j,n-1}}{{\cal Y}_j}-\frac{V_{k,n-1}}{{\cal Y}_k}\right)\;,~~~~
	\forall j\neq k\;.
	\label{Sn=}
	\xee
Next let us express $S_n$ using two different values of $k$,
say $k_1$ and $k_2$. Equating the two expressions and
simplifying the result, one obtains
	\xbe
	{\cal X}_j=\frac{{\cal Y}_j}{{\cal Y}_{k_2}}\left(
	{\cal X}_{k_2}+
	({\cal Y}_{k_1}{\cal X}_{k_2}-{\cal Y}_{k_2}{\cal X}_{k_1})
	\left[\frac{
	{\cal Y}_{k_2}V_{jn}-{\cal Y}_jV_{k_2n}}{
	{\cal Y}_{k_1}{\cal Y}_j V_{k_2n}-{\cal Y}_{k_2}{\cal Y}_jV_{k_1n}
	}\right]\right)\;,
	\label{Zj}
	\xee
Since $n$ does not appear in this equation except in the
square bracket, the content of the square bracket must be
independent of $n$. This argument may be used to determine
${\cal Y}_j$ by equating the square bracket on the right hand
side of (\ref{Zj}) with its value for $n=m_1$. This leads to 
	\xbe
	{\cal Y}_j={\cal Y}_{k_2} {\cal W}_{jn}\;,
	\label{Yj}
	\xee
where
	\xbe
	{\cal W}_{jn}:=\frac{
	( c V_{k_2m_1}-V_{k_1m_1})V_{jn}-
	( c V_{k_2n}-V_{k_1n})V_{jm_1}}{
	V_{k_2m_1}V_{k_1n}-V_{k_1m_1}V_{k_2n}}\;,	\label{Wj}
	\xee
and $c:={\cal Y}_{k_1}/{\cal Y}_{k_2}$.
Once again,  ${\cal W}_{jn}={\cal Y}_j/{\cal Y}_{k_2}$ must
be independent of $n$, i.e.,  for all $m$ and $n$, ${\cal W}_{jn}
={\cal W}_{jm}$. In particular ${\cal W}_{jn}={\cal W}_{jm_2}$,
where $m_2$ is some arbitrarily chosen fixed non-negative
integer. This equation may be used to express $V_{jn}$ in
terms of $V_{jm_1},~V_{jm_2},~V_{k_1n},~V_{k_2n}$, and
$c$ namely
	\xbea
	V_{jn}&=&\left[ \frac{cV_{k_2n}-V_{k_1n}}{cV_{k_2m_1}-V_{k_1m_1}}-
	\left(\frac{cV_{k_2m_2}-V_{k_1m_2}}{cV_{k_2m_1}-V_{k_1m_1}}\right)
	\left(\frac{V_{k_2m_1}V_{k_1n}-V_{k_1m_1}V_{k_2n}}{V_{k_2m_1}
	V_{k_1m_2}-V_{k_1m_1}V_{k_2m_2}}\right)\right]V_{jm_1}+\xnn\\
	&&\left[\frac{V_{k_2m_1}V_{k_1n}-V_{k_1m_1}V_{k_2n}}{V_{k_2m_1}
	V_{k_1m_2}-V_{k_1m_1}V_{k_2m_2}}\right] V_{jm_2}\;.
	\label{vjn-condi-0}
	\xeea
Next, consider the following two possibilities:
	\begin{itemize}
	\item[I)] $c\neq V_{k_1m_1}/V_{k_2m_1}$:\\
Then, the right-hand side of (\ref{vjn-condi-0}) does not actually
depend on $c$. In this case, one finds
	\[
	V_{jn}=\frac{
	(V_{k_1m_1}V_{jm_1}-V_{k_1m_2}V_{jm_2})V_{k_2n}+
	(V_{k_2m_2}V_{jm_2}-V_{k_2m_1}V_{jm_1})V_{k_1n}}{
	V_{k_1m_1}V_{k_2m_2}-V_{k_1m_2}V_{k_2m_1}}\;,\]
which is just Eq.~(\ref{vjn-condi-1}).
	\item[II)] $c=V_{k_1m_1}/V_{k_2m_1}$:\\
Then, according to the definition of $c$, i.e., $c:={\cal Y}_{k_1}/{\cal Y}_{k_2}$
and Eq.~(\ref{Sn=}) either all $S_n$ vanish --- this corresponds to the trivial
case $V_{jn}=0$ --- or  ${\cal X}_j/{\cal Y}_j={\cal X}_k/{\cal Y}_k$, i.e., the
ratio ${\cal X}_j/{\cal Y}_j$ does not depend on $j$. The latter implies ${\cal X}=
\eta {\cal Y}$ for some constant $\eta$. Moreover, in this case
$V_{jn}$ must satisfy
	\[V_{jn}=\frac{V_{jm_1}V_{k_1n}}{V_{k_1m_1}}\;,\]
which is just Eq.~(\ref{vjn-condi-2}). {\,\lower0.9pt\vbox{\hrule 
\hbox{\vrule height 0.2 cm \hskip 0.2 cm \vrule height 0.2 cm}\hrule}\,}
	\end{itemize}
	\end{itemize}
Note that in the latter case, Eq.~(\ref{vjn-condi-2}) together with
 ${\cal X}=\eta {\cal Y}$ and Eqs.~(\ref{Vjn}) and (\ref{V-s-=})
lead to 
	\xbea
	V_{jn}&=&v_n {\cal Y}_j\;,~~~~~v_n:=\eta(n+1)S_{n+1}+
	\sum_{m=0}^n(n-m+1)S_mS_{n-m}\;,
	\label{vjn-condi-3}\\
	V(x,t)&=&[\eta +\tilde S_+(x+t)]\tilde S'_+(x+t){\cal Y}(x-t)\;.
	\label{vjn-condi-4}
	\xeea
Now, substituting Eq.~(\ref{vjn-condi-4}) in Eq.~(\ref{s-=}), one can
easily show that indeed $\tilde S_+$ drops out of this equation and one
obtains $\tilde S_-=\eta$. Therefore the exact semiclassical wave
function $\psi=Re^{iS}$ is given by $R=R_-$ and $S=(\eta+\tilde S_+)
/R_-$. For example, consider choosing
	\begin{itemize}
	\item[5)] ${\cal Y}=\mu_-e^{\omega_-(x-t)}$ and 
	$(\eta+\tilde S_+)\tilde S_+'=\mu_+e^{\omega_+(x+t)}$ for
some constants $\mu_\pm$ and $\omega_\pm$:\\
Then, one has $V=\mu_+\mu_-e^{(\omega_++\omega_-)x}
e^{(\omega_+-\omega_-)t}$. In particular for $\omega_+=-\omega_-=:
\omega$, the potential depends only on $t$. In this case, one has
	\xbe
	V=\mu_+\mu_-e^{2\omega t}\;,~~~
	R=\sqrt{\frac{\nu_-+\mu_-e^{-\omega(x-t)}}{2\omega}}\;,~~~
	S=\pm2\sqrt{\frac{\mu_+e^{\omega x}+\nu_+e^{-\omega t}}{
	\mu_-e^{-\omega x}+\nu_-e^{-\omega t}}}\;,
	\label{v-r-s}
	\xee
where $\nu_\pm$ are also constants. The appearance of
the square roots in these equations is an indication that
for certain choices of $\mu_\pm$ and $\nu_\pm$ either $R$ or
$S$ can become imaginary in some regions of the Minkowski
space. Since $R$ and $S$ are assumed to be real, such a
semiclassical solution does not exist in these regions.
	\end{itemize}
This is another example of a case where the amplitude
of an exact semiclassical solution is not a slowly varying
function of its arguments.
 
A simple consequence of  Proposition~1 is:
	\begin{itemize}
	\item[] {\bf Corollary:} {\em The potentials of the form (\ref{V-s-=})
	which are analytic at $(x=0,t=0)$ form a proper subset of the set
	of all potentials, i.e., for an arbitrary potential which is analytic at
	$(x=0,t=0)$, an exact semiclassical solution of the $(1+1)$-dimensional
	Klein-Gordon  equation (\ref{kg-eq}), with $R_+=0$ and
	$A_\mu=0$, may not exist.}
	\end{itemize}

Eqs.~(\ref{vjn-condi-1}) and (\ref{vjn-condi-2}) provide
a useful criterion for finding out whether a given potential allows for
an exact semiclassical solution with $R_+=0$ or not. If the result is
positive, then Eqs.~(\ref{Sn=}), (\ref{Zj}), and (\ref{Yj}), may be used
to determine $R_-$ and $\tilde S_+$. These are then used  to integrate
Eq.~(\ref{s-=}) which yields $\tilde S_-$ and consequently the wave
function $\psi=R_-\exp[i(\tilde S_-+\tilde S_+)/R_-]$.

Proposition~1 only applies to the cases where $R_+=0$.  One might try
to employ a similar method to treat the more general case, where $R_+$
is also an undetermined function. For this purpose, one must first write
Eq.~(\ref{1-1-V}) as a polynomial equation in $V$, $R_\pm$, $\tilde S_\pm$ and
their derivatives and substitute the power series expansions of these
functions in the resulting expression.  This leads to an infinite system of very
complicated coupled nonlinear algebraic equations for the coefficients
of $R_\pm$, $\tilde S_\pm$ whose analytic solution has not been possible.
Although a similar proof is lacking for the most general case, further
inspection of Eq.~(\ref{1-1-V}) suggests that this equation also restricts
the form of the potential. Hence, in general  for an arbitrary potential
an exact semiclassical solution of the Klein-Gordon equation does not
exist.

\section{Exact Semiclassical Wave Functions of the Universe}

Consider the  Wheeler-DeWitt equation for the closed FRW cosmological
model coupled to a scalar field $\phi$, \cite{page,wiltshire},
	\xbe
	\left[-\partial_\alpha^2+\partial_\phi^2+
	e^{4\alpha}-e^{6\alpha}{\cal V}(\phi)\right]\psi=0\;,
	\label{wdw}
	\xee
where $\alpha:=\ln a$, $a$ is the scale factor of the FRW model,
${\cal V}$ is the matter field potential. Here, the cosmological constant
is assumed to vanish, the usual factor ordering prescription
\cite{page,wiltshire} is adopted, and the natural units in which
the Planck mass is set to unity is used. Clearly, this is a Klein-Gordon
equation in $(1+1)$-dimensional Minkowski (minisuper)space with
potential 
	\xbe
	V=-e^{4\alpha}+e^{6\alpha}{\cal V}(\phi)\;,
	\label{wdw-v}
	\xee
where $(\alpha,\phi)$ play the role of $(t,x)$.

The solutions of Eq.~(\ref{wdw}) have been studied mostly for the massless
(${\cal V}=0$) and massive (${\cal V}=m^2\phi^2$) scalar fields in the
literature \cite{approx,kiefer-88,h-p-90,page,kim-92,wiltshire,p-wdw}. This is done
by making use of different approximation schemes except for the rather trivial
and much less interesting massless case for which the exact solution is known
\cite{massless,h-p-90}. The approximate solutions of Eq.~(\ref{wdw}) are
usually developed by making particular assumptions for the boundary
conditions, semiclassicality of the solution, or restricting to particular regions
of the minisuperspace in which the Wheeler-DeWitt equation (\ref{wdw})
simplifies.

In view of the developments reported in the preceding sections,
the assumption of the exactness of the semiclassical approximation
provides a direct link between the choice of the boundary conditions
and  the form of the potential.  This is done through the function
${\cal F}$ defined by (\ref{1-1-V}) which can be viewed as a function
from the set of boundary conditions to the set of  potentials $V$.
In section~4, I have shown for the case $R_+=0$ that this
function is not onto, i.e., there are potentials which do not admit exact
semiclassical solutions.  Here, however, one is interested in a special
class of potentials, namely those of the form (\ref{wdw-v}).  Hence, the
relevant problem is to find the intersection ${\cal U}$ of the image of
${\cal F}$ and the set of potentials of the form (\ref{wdw-v}). One can easily
show that ${\cal U}$ is not empty. For example the massless case,
where $V$ is an exponential function of $\alpha$, can be easily put in
the form (\ref{1-1-V}), i.e., it belongs to ${\cal U}$. In fact, two possible
choices for $R_\pm$ and $\tilde S_\pm$ which lead to this potential
were already given in examples~3 and~5 of section~4
(with $\omega=2$). The following are nontrivial examples of the
potentials of the form (\ref{wdw-v}) which also belong to ${\cal U}$.
They are obtained by setting $R_+=0$ and making simple choices for
the functions ${\cal X}$, ${\cal Y}$, and $\tilde S_+$ of Eq.~(\ref{V-s-=})
so that Eq.~(\ref{wdw-v}) is also satisfied. This together with Eq.~(\ref{s-=})
yield $\tilde S_-$.
	\begin{itemize}
	\item[1)] {${\cal V}=\lambda e^{2\phi}$ with $\lambda>0$}\\
In this case, the choices ${\cal X}=-e^{-2(\phi-\alpha)}$, $R_-=2e^{\phi-\alpha}/
\sqrt{\lambda}$, and  $\tilde S_+=e^{2(\phi+\alpha)}/2$ satisfy
Eq.~(\ref{V-s-=}). The solution of (\ref{s-=}) then leads to $\tilde S_-=
ce^{\phi-\alpha}+1/\lambda$, where $c$ is a constant. The exact
semiclassical solution is given by
	\[ R=\frac{2e^{\phi-\alpha}}{\sqrt{\lambda}}\;,~~~~
	S=\sqrt{\lambda}(e^{\phi+3\alpha}+\frac{2e^{\alpha-\phi}}{
	\lambda}+2c)\;;\]
	\item[2)] {${\cal V}=\lambda e^{-4\phi}$ with arbitrary $\lambda$}\\
In this case, one has ${\cal X}=\lambda e^{-5(\phi-\alpha)}/c_1$,
$R_-=2c_1[c_2-e^{-2(\phi-\alpha)}]^{-1/2},~\tilde  S_+=c_1e^{\phi+\alpha}$,
and 
	\[ \tilde S_-=-\frac{\lambda c_1}{8}\left[2e^{-3(\phi-\alpha)}+
	3c_2 e^{-(\phi-\alpha)}+3c_2^2e^{\phi-\alpha}\left(
	\frac{\tan^{-1}\sqrt{c_2 e^{2(\phi-\alpha)}-1}+c_3}{
	\sqrt{c_2 	e^{2(\phi-\alpha)}-1}}\right)\right]\;,\]
where $c_1, ~c_2$ and $c_3$ are constants. The exact
semiclassical solution is therefore given by
	\xbea
	 R&=&\frac{2c_1}{\sqrt{c_2-e^{-2(\phi-\alpha)}}}\;,\xnn\\
	S&=&\left( \frac{e^{\phi+\alpha}}{2}-\frac{\lambda
	\left[2e^{-3(\phi-\alpha)}+3c_2 e^{-(\phi-\alpha)}\right]}{16}
	\right)
	\sqrt{c_2-e^{-2(\phi-\alpha)}}	-\frac{3\lambda c_2^2}{16}\left(
	\tan^{-1}\sqrt{c_2 e^{2(\phi-\alpha)}-1}+c_3 \right)	\;.\xnn
	\label{sol-2}
	\xeea
Note that in this case, there is always a region of the
minisuperspace defined by  $e^{\phi-\alpha}<c_2^{-1/2}$
where a semiclassical solution does not exist.
	\end{itemize}

The next logical step is to explore the existence of exact semiclassical
solutions of the Wheeler-DeWitt equation with matter potentials of the
form ${\cal V}=\lambda \phi^{2p}$. These are among the potentials which
lead to inflationary classical solutions, \cite{inflation}. The simplest case
is that of a massive scalar field, i.e., $\lambda=m^2,~p=1$. In the remainder
of this section, I shall restrict to the semiclassical solutions with $R_+=0$.
The existence of this type of solutions can be easily decided using
Eqs.~(\ref{vjn-condi-1}) and (\ref{vjn-condi-2}). One simply needs to
compute the coefficients $V_{jn}$ of (\ref{v-exp}) and check whether
they satisfy one of these equations. A simple calculation shows that
for $p=1,~2,~3$, none of these equations are satisfied. Hence, the
matter potentials $\lambda\phi^2$,
$\lambda\phi^4$, and $\lambda\phi^6$ do not admit exact `right-going'
($R_+=0$) semiclassical solutions. This is done by choosing the
integers $j,~n,~k_1~,k_2~,m_1,$ and $m_2$ in such a way that
both Eqs.~(\ref{vjn-condi-1}) and (\ref{vjn-condi-2}) fail. In order
to demonstrate this, let us denote the right-hand sides of
Eqs.~(\ref{vjn-condi-1}) and (\ref{vjn-condi-2}) by $V^{(1)}_{jn}$
and $V^{(2)}_{jn}$, respectively. Then,
	\begin{itemize}
	\item[---]  for $p=1$ and $k_1=m_1=0,~k_2=m_2=1$, one finds
	\[V_{22}=-4+\frac{9\lambda}{4}\;,~~~
	V^{(1)}_{22}=-4+\frac{13\lambda}{8}-\frac{\lambda^2}{16}\;,~~~
	V^{(2)}_{22}=-(2-\frac{\lambda}{4})^2\;,\]
	\[V_{32}=\frac{8}{3}+\frac{9\lambda}{4}\;,~~~
	V^{(1)}_{32}=\frac{8}{3}-\frac{23\lambda}{12}-\frac{3\lambda^2}{16}\;,~~~
	V^{(2)}_{32}=\frac{8}{3}-\frac{11\lambda}{6}+\frac{3\lambda^2}{16}\;.\]
	\item[---]  for $p=2$ and $k_1=m_1=0,~k_2=m_2=2$, one finds
	\[V_{31}=\frac{8}{3}+\frac{\lambda}{4}\;,~~~
	V^{(1)}_{31}=V^{(2)}_{31}=\frac{8}{3}\;,\]
	\[V_{33}=\frac{16}{9}-\frac{9\lambda}{8}\;,~~~
	V^{(1)}_{33}=\frac{16}{9}+\frac{3\lambda}{8}\;,~~~
	V^{(2)}_{33}=\frac{16}{9}\;.\]
	\item[---]  for $p=3$ and $k_1=m_1=0,~k_2=m_2=3$, one finds
	\[V_{24}=-\frac{4}{3}+\frac{15\lambda}{64}\;,~~~V^{(1)}_{24}=
	V^{(2)}_{24}=-\frac{4}{3}\;.\]
	\end{itemize}
Note that for $\lambda=0$, i.e., the massless case, the values
of $V_{jn}$ in the above list match the values of $V^{(1)}_{jn}$
and $V^{(2)}_{jn}$. This is in agreement with the fact that for the
massless case one does in fact have right-going exact semiclassical
solutions. Eqs.~(\ref{v-r-s}) with $(t,x)\to(\alpha,\phi)$, $\omega=2$,
and $\mu_-=-1/\mu_+$, provide a concrete example.

Furthermore, one knows from the studies of the inflationary
cosmological models that for the polynomial matter potentials
$\lambda\phi^{2p}$ the coupling constant $\lambda$ must be a very
small number. For example for the massive case, where $p=1$ and
$\lambda=m^2$, these theories predict $m\approx10^{-6}$, i.e.,
$\lambda\approx10^{-12}$, \cite{inflation}. Thus, although there is no
right-going semiclassical solutions for $p=1,2,3$, at least for small
values of $u=\phi-\alpha$ and $v=\phi+\alpha$, where one can
neglect forth and higher order terms in the power series expansion
of $V$, the semiclassical approximation seems to be reliable. 

The phrase {\em semiclassical approximation} is used in quantum
cosmology in a very crude way. One usually makes additional
assumptions such as the adiabaticity of the evolution \cite{kiefer-88,conradi}
to reduce the situation to the one-dimensional quantum
mechanical case. In this way, one is able to express the condition of the
validity of the semiclassical approximation as a simple limitation on the
range of values of the matter potential, \cite{wiltshire}. The consistency of
these assumptions with the validity of the semiclassical approximation
is either left unchecked or a set of conditions is imposed which render
the scheme consistent. These conditions are usually sufficient
conditions not necessary.  Hence, in general they may be too
restrictive. The situation is very similar to restricting the exact
semiclassical wave functions to those with slowly varying amplitudes.
As shown in the preceding sections, this is an absolutely unnecessary
restriction. The approach pursued in the article also allows for a
precise definition of a more general semiclassical approximation
where the solution to the dynamical equations is approximated with
the general semiclassical wave functions introduced in section~1.
This will be discussed next. 

\section{Semiclassical Perturbation Theory}

Let us first define a {\em semiclassical potential} $V_0$ to be a potential
which corresponds to an exact semiclassical solution of the dynamical
equation. In view of the results of sections~2 and 3, the set 
of semiclassical potentials is in fact much larger than one usually expects.
This suggests a generalized notion of  {\em semiclassical expansion}
and in particular {\em semiclassical  approximation} which corresponds
to a perturbation theory around the semiclassical potentials.

Let $V$ be an arbitrary potential, $\psi=Re^{iS/\hbar}$ be the solution
of the dynamical equation, and $\epsilon\in\xR$ be a perturbation
parameter. Then, the semiclassical perturbation theory corresponds to
	\xbea
	V&=&V_0+\delta V\;,~~~~\delta V=\epsilon~V_{\rm p}
	\label{pert-v}\\
	R&=&R_0+\delta R\;,~~~~\delta R=\sum_{\ell=1}^\infty
	\epsilon^\ell R_\ell\;,
	\label{pert-r}\\
	S&=&S_0+\delta S\;,~~~~\delta S=\sum_{\ell=1}^\infty
	\epsilon^\ell S_\ell\;,
	\label{pert-s}
	\xeea
where $V_0$ and $\psi_0=R_0 e^{iS_0/\hbar}$ are the
semiclassical potential and wave function defined by the
boundary conditions of the problem. Substituting
Eqs.~(\ref{pert-v}) -- (\ref{pert-s}) in the dynamical equations
and treating $\epsilon$ as an independent parameter, one
obtains an infinite family of equations which can be iteratively
solved to yield $R_\ell$ and $S_\ell$. In particular, the equations
obtained in the $\ell$-th order are two linear coupled differential
equations in $R_\ell$ and $S_\ell$ with vanishing boundary
conditions. This is because the original boundary conditions
are already imposed in the determination of $R_0$ and $S_0$.

Eqs.~(\ref{pert-r}) and (\ref{pert-s}) yield what one might call
a {\em semiclassical perturbation expansion}. Note that
they are not power series in $\hbar$ but in the perturbation
parameter $\epsilon$. The zero-th order terms correspond to
the semi-classical wave function. Thus, the {\em semiclassical
approximation} is defined by $R\approx R_0$ and $S\approx
S_0$. Note also that the semiclassical wave function $\psi_0$ 
and potential $V_0$ are uniquely determined by the boundary
conditions. The choice of the perturbation parameter is, however,
made by the physics of the problem. A typical example of a
perturbation parameter is the coupling constant $\lambda$
of the preceding section.

Let us next list the equations governing the first and second
order terms in the semiclassical perturbation expansion.
	\begin{itemize}
	\item[---]  {\em Nonrelativistic QM: Schr\"odinger
	Equation}
	\begin{itemize}
	\item[] First order  (post-semiclassical) corrections, i.e., 
	equations determining $R_1$ and $S_1$:
	\xbea
	\partial_t S_1+\frac{1}{m}({\bf\nabla}S_0-{\bf A})\cdot
	{\bf\nabla}S_1+V_{\rm p}-\frac{\hbar^2{\bf\nabla}^2R_1}{2mR_0}
	&=&0\;,\xnn
	\\
	2\partial_t (R_0R_1)+\frac{1}{m}{\bf\nabla}\cdot\left[
	R_0^2{\bf\nabla }S_1+2R_0R_1({\bf\nabla}S_0-{\bf A})\right]
	&=&0\;.\xnn
	\xeea
	\item[] Second order corrections, i.e., equations determining
	$R_2$ and $S_2$:
	\xbea
	\partial_t S_2+\frac{1}{2m}\left[({\bf\nabla}S_1)^2+
	2({\bf\nabla}S_0-{\bf A})\cdot {\bf\nabla}S_2\right]-
	\frac{1}{2m}\left(\frac{{\bf\nabla}^2R_2}{R_0}-
	\frac{R_1{\bf\nabla}^2R_1}{R_0^2}\right)&=&0\;,
	\xnn\\
	2\partial_t (R_0R_2)+\partial_t R^2_1+\frac{1}{m}{\bf\nabla}
	\cdot\left[(R_1^2+2R_0R_2)({\bf\nabla}S_0-{\bf A})+
	2R_0R_1{\bf\nabla}S_1+R_0^2{\bf\nabla}S_2\right]&=&0\;.\xnn
	\xeea
	\end{itemize}
	These equations are obtained by substituting Eqs.~(\ref{pert-v}) --
	(\ref{pert-s}) in Eqs.~(\ref{q-hj-eq}) and (\ref{conti}).
	\item[---]  {\em Relativistic QM: Klein-Gordon
	Equation} ($c=\hbar=1$)
	\begin{itemize}
	\item[] First order  (post-semiclassical) corrections, i.e., 
	equations determining $R_1$ and $S_1$:
	\xbea	
	2(\partial^\mu S_0-A^\mu)\partial_\mu S_1+V_{\rm p}+
	\frac{\partial^\mu\partial_\mu R_1}{R_0}&=&0\;,
	\label{l=1-1/kg}\\
	\partial_\mu\left[2R_0R_1(\partial^\mu S_0-A^\mu)+
	R_0^2\partial^\mu S_1\right]&=&0\;.
	\label{l=1-2/kg}
	\xeea
	\item[] Second order corrections, i.e., equations determining
	$R_2$ and $S_2$:
	\xbea
	2(\partial^\mu S_0-A^\mu)\partial_\mu S_2+\partial^\mu S_1\partial_\mu S_1+
	\frac{\partial^\mu\partial_\mu R_2}{R_0}-
	\frac{R_1\partial^\mu\partial_\mu R_1}{R_0^2}&=&0\;,\xnn\\
	\partial_\mu\left[R_0^2\partial^\mu S_2+2R_0R_1\partial^\mu S_1+
	 (R_1^2+2R_0R_2)(\partial^\mu S_0-A^\mu)\right]&=&0\;.\xnn
	\xeea
	\end{itemize}
	These equations are obtained by substituting Eqs.~(\ref{pert-v}) --
	(\ref{pert-s}) in Eqs.~(\ref{q-hj-eq/kg}) and (\ref{conti/kg}). They
	can be further simplified. For example, consider the first order
	equations~(\ref{l=1-1/kg}) and (\ref{l=1-2/kg}), and define 	$p_0^\mu:=\partial^\mu S_0-A^\mu$ and $T_{\pm1}:=R_1\pm R_0
	S_1$. Then using the fact that $R_0$ and $S_0$ define
	a semiclassical wave function, i.e., $\partial_\mu\partial^\mu
	R_0=0$ and $\partial_\mu(R_0^2 p_0^\mu)=0$, one can show
	that  Eqs.~(\ref{l=1-1/kg}) and (\ref{l=1-2/kg}) are equivalent 
	to
	\xbe
	\left[\partial_\mu\partial^\mu \pm 2p_0^\mu(\partial_\mu-
	\partial_\mu\ln R_0)\right] T_{\pm 1}=-R_0V_{\rm P}\;.
	\label{T}
	\xee
	These are two separate equations for $T_{\pm 1}$ whose
	solution yields $R_1$ and $S_1$.
	\end{itemize}

These equations appear to be even more difficult to solve 
than the original dynamical equations. Note, however, that they
are to be solved with vanishing boundary conditions. The advantage
of this scheme is that the information on the boundary conditions of the
original dynamical equation is restored in the definition of $R_0$ and
$S_0$. The higher order corrections are affected by these boundary
conditions only through $R_0$, $S_0$ and the perturbation potential
$V_{\rm p}$ which appears in the first order of perturbation. In this sense,
the semiclassical perturbation theory has a universal character.

As a concrete example consider the Wheeler-DeWitt equation
of section~5 with a matter potential of the form
${\cal V}=\lambda f(\phi)$, where $f$ is some real-valued function.
Let the boundary conditions be such that one
recovers example~5 of section~4 with $\omega=2$,
$\mu_\pm=\mp1$,  $\nu_-=0$, and $\nu_+=\nu$. Then replacing $(t,x)$ with
$(\alpha,\phi)$, $(V,R,S)$ with $(V_0,R_0,S_0)$, and adopting the
positive sign in Eqs.~(\ref{v-r-s}), one has
	\xbe
	V_0=-e^{4\alpha}\;,~~~~R_0=\frac{1}{2}
	e^{-(\phi-\alpha)}\:,~~~~
	S_0=2\sqrt{\nu e^{2(\phi-\alpha)}-e^{4\phi}}\;.
	\label{v-r-s-0}
	\xee
These correspond to an approximate semiclassical solution
which is an exact solution of the massless case. Clearly, the
perturbation potential $V_{\rm P}$ is given by $e^{6\alpha}f(\phi)$
and the perturbation parameter is $\lambda$. The first order
(post-semiclassical) correction $(R_1,S_1)$ to $(R_0,S_0)$ is
obtained by solving Eqs.~(\ref{T}) which take the form
	\xbe
	\left\{
	-\partial_\alpha^2+\partial_\phi^2\pm[\xi (\partial_\alpha-1)
	+\zeta(\partial_\phi+1)]\right\}T_{\pm 1}=-\frac{1}{2}
	e^{-\phi+7\alpha}f(\phi)\;,
	\label{T'}
	\xee
with
	\[ \xi:=-2\partial_\alpha S_0=\frac{4\nu e^{2(\phi-\alpha)}}{
	\sqrt{\nu  e^{2(\phi-\alpha)}-e^{4\phi}}}\:,~~~~
	\zeta:=2\partial_\phi S_0=\frac{4(\nu e^{2(\phi-\alpha)}-
	2e^{4\phi})}{\sqrt{\nu e^{2(\phi-\alpha)}-e^{4\phi}}}\:.\]
Although solving Eqs.~(\ref{T'}) seems much more difficult
than solving the original Wheeler-DeWitt equation (\ref{wdw}),
one must recall that these equations are to be solved with
vanishing boundary conditions. In this way one can
at least devise an efficient numerical scheme which
can treat the solution of these and the equations
for the higher order corrections for arbitrary matter
potentials. 

In general, such a scheme should first compute the
semiclassical potential $V_0$ and wave function
$R_0e^{iS_0}$ using the boundary conditions. This
would yield the perturbation potential $V_{\rm p}$
(after the perturbation parameter is identified according
to the physical characteristics of the problem). Then,
it should numerically integrate the equations satisfied
by $R_\ell$ and $S_\ell$ with vanishing boundary
conditions.

Finally, let me emphasize that for the Wheeler-DeWitt  equation
with the massive scalar field. The perturbation parameter is already
extremely small $\lambda=m^2={\cal O}(10^{-12})$, therefore the first
order corrections provide solutions which are valid up to the order
$\lambda^2={\cal O}(10^{-24})$. This suggests that the domain of 
the validity of the first order perturbation theory is indeed quite
large.
 
\section{Conclusion}

In this article I have tried to demonstrate how the simple observation
that the traditional condition of validity of the semiclassical
approximation is only a sufficient condition, can be used to
introduce the notions of exact semiclassical wave function
and potential. 

I have shown that the semiclassical wave function
and potential are both determined by the boundary conditions
of the problem uniquely. I have also given the full classification
of exact semiclassical wave functions and potentials for
the Schr\"odinger and Klein-Gordon equations in arbitrary
dimensions. The analysis of the one-dimensional Schr\"odinger
equation is much simpler than the multi-dimensional case. 
For the Klein-Gordon equation in (1+1)-dimensions which is
directly relevant to the solution of the Wheeler-DeWitt
equation for FRW scalar field minisuperspace models,
I have explicitly constructed semiclassical wave functions
and potentials. For this case, I have also developed a practical
criterion for checking whether a given potential allows for a
right-going exact semiclassical solution. I have then used
this criterion to study the semiclassical solutions of the minisuperspace
Wheeler-DeWitt equation. For the polynomial matter potentials
of the form $\lambda\phi^{2p}$ with $p=1,2,3$, I have shown that
a semiclassical solution of the Wheeler-DeWitt equation does
not exist. However, the non-existence proof relies on the
non-zero value of the coupling constant $\lambda$ which is
expected to be an extremely small number. This motivated the
development of a semiclassical perturbation theory which
yields the semiclassical approximation as the zero-th order
term in the semiclassical perturbation expansion. The higher
order terms satisfy coupled linear differential equations with
vanishing boundary conditions. 

The resulting semiclassical approximation and the domain
of its reliability are different from the traditional semiclassical
approximation. They coincide for the cases where the semiclassical
approximation is exact, i.e., the potential and wave function are
semiclassical. The semiclassical expansion developed in this
paper is a perturbation expansion about the exact semiclassical
potential defined by the boundary conditions. This is in contrast
with the traditional semiclassical expansion which is an $\hbar$ or
$M_p^{-1}$-expansion of the solution of the dynamical equation.

\section*{Acknowledgements}

I would like to thank  B.~Darian and M.~Razavi for helpful
discussions. I would also like to acknowledge the financial
support of the Killam Foundation of Canada.


\end{document}